\documentclass[conference]{IEEEtran}
\IEEEoverridecommandlockouts
\usepackage{cite}
\usepackage{amsmath,amssymb,amsfonts}
\usepackage{algorithmic}
\usepackage{graphicx}
\usepackage{textcomp}
\usepackage{enumitem}
\usepackage{booktabs}
\usepackage[utf8]{inputenc}
\usepackage{balance}
\usepackage[table,xcdraw]{xcolor}
\usepackage{amsmath}
\usepackage{adjustbox}
\usepackage{xcolor}
\usepackage{caption}

\usepackage[T1]{fontenc} 
\usepackage{pythonhighlight}
\usepackage{array}
\usepackage{tikz}
\usepackage[most]{tcolorbox}

\usepackage{xcolor}
\usepackage{subcaption}
\def\BibTeX{{\rm B\kern-.05em{\sc i\kern-.025em b}\kern-.08em
    T\kern-.1667em\lower.7ex\hbox{E}\kern-.125emX}}

\usepackage{siunitx}
\usepackage{pifont}

\usepackage{float}   
\usepackage{minted}  

\usepackage{listings}
\usepackage{xcolor}

\lstdefinelanguage{Rego}{
  morekeywords={package, import, default, not, true, false, null},
  sensitive=true,
  comment=[l]{\#},
  morecomment=[s]{/*}{*/},
  morestring=[b]"
}

\lstdefinestyle{regostyle}{
  basicstyle=\ttfamily\footnotesize,
  keywordstyle=\color{blue!70!black}\bfseries,
  commentstyle=\color{gray!70},
  stringstyle=\color{green!50!black},
  numbers=left,
  numberstyle=\tiny\color{gray},
  stepnumber=1,
  numbersep=8pt,
  frame=lines,
  framerule=0.5pt,
  framesep=2mm,
  breaklines=true
}

\lstdefinelanguage{bash}{
  morekeywords={sudo,apt,opa,eval,cd,ls,cat,echo,export},
  sensitive=true,
  morecomment=[l]{\#},
  morestring=[b]",
  morestring=[b]'
}

\lstdefinestyle{termstyle}{
  basicstyle=\ttfamily\footnotesize,
  keywordstyle=\bfseries\color{blue!70!black},
  commentstyle=\itshape\color{gray!70},
  stringstyle=\color{green!50!black},
  numbers=left,
  numberstyle=\tiny\color{gray},
  frame=lines,
  framerule=0.5pt,
  framesep=2mm,
  breaklines=true,
  showstringspaces=false,
  columns=fullflexible
}
\usepackage{hyperref}
\begin{document}

\title{Empirical Characterization of Logging Smells in Machine Learning Code.
}
\author{
Patrick Loic Foalem\textsuperscript{\dag}, 
Leuson Da Silva\textsuperscript{\dag}, 
Foutse Khomh\textsuperscript{\dag}, 
Ettore Merlo\textsuperscript{\dag}, 
Heng Li\textsuperscript{\dag}\\
\{patrick-loic.foalem, leuson-mario-pedro.da-silva, foutse.khomh, ettore.merlo, heng.li\}@polymtl.ca\\
\textsuperscript{\dag}Polytechnique Montr\'eal, Montr\'eal, Quebec, Canada
}

\maketitle

\begin{abstract}
\underline{Context:} Logging is a fundamental yet complex practice in software engineering, essential for monitoring, debugging, and auditing software systems. With the increasing integration of machine learning (ML) components into software systems, effective logging has become critical to ensure reproducibility, traceability, and observability throughout model training and deployment. Although various general-purpose and ML-specific logging frameworks exist, little is known about how these tools are actually used in practice or whether ML practitioners adopt consistent and effective logging strategies. To date, no empirical study has systematically characterized recurring bad logging practices--or logging smells--in ML System.
\underline{Goal:} This study aims to empirically identify and characterize logging smells in ML systems, providing an evidence-based understanding of how logging is implemented and challenged in practice.
\underline{Method:} We propose to conduct a large-scale mining of open-source ML repositories hosted on GitHub to catalogue recurring logging smells. Subsequently, a practitioner survey involving ML engineers will be conducted to assess the perceived relevance, severity, and frequency of the identified smells.
\underline{Limitations:} 
While our findings may not be generalizable to closed-source industrial projects, we believe our study provides an essential step toward understanding and improving logging practices in ML development.
\end{abstract}


\section{Introduction}
\label{sec:introduction}
Logging is a fundamental engineering practice that supports the monitoring, debugging, and auditing of software systems \cite{foalem2025logging, foalem2024studying, yuan2012characterizing, fu2014developers}. 
However, integrating logging effectively into source code remains a challenging task \cite{li2020qualitative}. Developers must decide on the appropriate logging levels, determine which variables or events to log, and configure logging frameworks consistently across different environments and libraries \cite{li2020qualitative, batoun2024literature}. 

As machine learning (ML) models are increasingly deployed in complex, real-world environments, the need for effective and systematic logging in ML workflows has become critical \cite{foalem2024studying}. 
Logging enables ML practitioners to capture key information during training, evaluation, and deployment, thereby maintaining a detailed record of model behavior, performance, and experimental conditions. 
In practice, two broad categories of logging libraries are employed in ML development: 
(1) \textit{general-purpose logging libraries} (e.g., the \texttt{logging} module in Python\footnote{https://docs.python.org/3/library/logging.html}) that are traditionally used in traditional software systems, and 
(2) \textit{ML-specific logging frameworks} (e.g., MLflow,\footnote{https://mlflow.org/} Weights \& Biases\footnote{https://wandb.ai/site/}) designed to track experiments, metrics, parameters, and artifacts in ML components \cite{foalem2024studying}. Unlike general-purpose logging libraries that primarily record textual events and system states, ML-specific logging frameworks provide dedicated support for experiment tracking--such as structured metric logging, artifact versioning, hyperparameter recording, and model lifecycle management--which are essential for ensuring reproducibility in ML workflows. 
In contrast to other traditional systems, ML workflows introduce unique sources of variability--such as stochastic training behavior, dataset-dependent execution paths, and evolving model artifacts--that require specialized logging support, suggesting that the resulting taxonomy may capture ML-specific smells not observed in traditional logging environments.

While traditional software applications typically rely on a single, general-purpose logging library, recent empirical studies have begun to catalog common logging issues in such contexts \cite{saarimaki2024taxonomy}. In contrast, despite the growing use of logging tools in ML-based systems, there has been no systematic investigation of the recurring bad practices or smells in ML logging. 
This represents a significant knowledge gap at the intersection of software engineering and machine learning practice.

To address this gap, our study aims to provide a catalogue of logging smells in ML code. 
We propose to mine a large sample of open-source ML repositories to identify recurring logging smells and then conduct a follow-up survey of ML practitioners to assess the prevalence, perceived severity, and relevance of these practices in real-world settings. 
By combining large-scale code analysis with practitioner perspectives, our contribution will deepen the understanding of ML logging practices, while supporting the development of guidelines and tooling for more effective logging and observability in ML systems.

\section{Background and Related Work}
Saarim{\"a}ki et al. \cite{saarimaki2024taxonomy} propose a comprehensive taxonomy of log smells grounded in traditional software systems. Their work identifies recurring design and implementation deficiencies in logging statements, like inconsistent log levels, missing contextual data, and verbose or redundant logs. 
In our proposed study, we aim to extend the previous taxonomy by focusing on the domain of machine learning (ML) systems, where logging practices exhibit new forms of challenges due to the complexity of ML pipelines, dynamic training configurations, and framework heterogeneity (e.g., TensorFlow, PyTorch). 
This way, the taxonomy by Saarim{\"a}ki et al. serves as a conceptual precursor, providing a classification lens that this research empirically adapts and validates in ML-specific contexts. 
As a result, we expect to bridge the gap between traditional logging research and the emerging challenges of observability in ML-based software systems.

In the same direction, Washizaki et al. \cite{washizaki2019studying} present an empirical investigation into logging behavior in traditional software systems. 
For that, the authors explore how developers insert, structure, and maintain log statements over the evolution of software. While its focus is on general-purpose software (traditional software development), it offers valuable insights into patterns of logging usage, evolution, and maintenance-related issues in non-ML contexts. 
Compared to our work, we aim to adapt and extend this line of investigation by focusing specifically on machine learning (ML) codebases, covering their main particularities, like data pipelines, model evaluation, and stochastic behaviors that are not typical in traditional software. 
As a result, logging smells and challenges in ML systems are expected to diverge from those observed in traditional systems. Thus, we can explore whether its logging usage patterns hold in ML settings, identify divergence, and propose ML-centric logging guidelines that complement the broader observations from that earlier work.

While our study aims to characterize recurring logging smells and assess their relevance to practitioners, Chen et al. \cite{chen2025empirical} provide a domain-specific empirical characterization of how logging is performed in GPU-accelerated deep learning (DL) projects. 
Their analysis of 33 CUDA-based open-source repositories reveals that most logging statements occur during the model training, model loading, and evaluation/validation phases, with logs primarily serving monitoring and model-related tracking purposes. 
The authors further distinguish between general-purpose logging frameworks (e.g., logging, loguru) and DL-specific ones (e.g., TensorBoard, MLflow, WandB), showing that developers typically employ both in combination to address different observability needs.
Compared to our proposed study, the authors provide empirical evidence on where and why logging occurs in DL systems, but do not examine the quality or problematic nature of those practices. 
Our study extends beyond their findings by investigating smells, thus focusing on the effectiveness rather than the frequency or phase distribution of logging.

Rodriguez et al. \cite{rodriguez2025automated} investigated the capacity of GPT-4o-mini to automatically generate file-level logging statements for ML applications, showing that while LLMs can reproduce human logging positions with moderate accuracy, they also introduce substantial overlogging and often misalign with project-specific conventions. Their findings highlight both the promise and limitations of LLM-assisted logging. While their study focuses on automated log generation, our work is complementary: we aim to empirically characterize recurring logging smells in ML codebases and understand developer perceptions of these smells. Together, these lines of research underscore the need for a systematic understanding of logging practices and quality issues in ML systems.

\section{Goal of the Study}
The primary goal of this study is to empirically investigate how logging practices are implemented in machine learning (ML)-based systems and to identify recurring bad practices, or \textit{smells}, that emerge in such contexts. 
Specifically, this study explores whether the coexistence of two distinct categories of logging libraries--general-purpose logging libraries (e.g., \texttt{logging} for Python) and ML-specific logging frameworks (e.g., MLflow, Weights \& Biases)--introduces new forms of complexity that can lead to logging smells. 
The study aims to build a comprehensive understanding of how ML practitioners use these libraries in open-source projects, and how such practices impact observability, reproducibility, and experiment tracking.

In pursuit of this goal, the investigation follows two main directions:  
(i) identifying and cataloging logging smells through large-scale mining of open-source ML repositories, and (ii) evaluating the perceived relevance and severity of these smells through a survey with ML practitioners.  

As the \textbf{leading hypothesis} of this study, we have:
\begin{quote}
    \textit{Since ML-based systems rely on both general-purpose and ML-specific logging libraries to capture runtime behavior and experiment metadata, the simultaneous use of these heterogeneous logging mechanisms may introduce redundancy, inconsistency, and fragmentation in the logged information, while also giving rise to novel ML-specific logging smells related to experiment tracking, stochastic training behavior, and model lifecycle management.}
\end{quote}
This hypothesis is grounded in prior empirical evidence showing that logging practices in traditional software systems already suffer from inconsistencies and maintenance challenges \cite{saarimaki2024taxonomy}. However, ML development introduces \emph{additional and fundamentally different} sources of complexity that are not typically present in other heterogeneous environments. ML workflows rely on experiment tracking tools, structured metric logging, artifact versioning, stochastic training behavior, and data-dependent execution paths, all of which interact with general-purpose logging in ways that can produce new forms of redundancy, inconsistency, or omission. Because these ML-specific activities generate information that must be logged across multiple abstraction layers (data, model, training loop, hardware), we expect logging flaws to manifest differently than in other traditional software logging scenarios, thereby justifying the search for distinct ML-specific smells.

\section{Research Questions and hypotheses}
\label{sec:RQ}
As previously reported, out mains goal is to empirically identify and characterize logging smells in ML systems, and to understand their perceived relevance from the perspective of ML practitioners. 
To achieve our goal, we define two main research questions, each accompanied by corresponding hypotheses.

\subsection*{RQ1: What logging smells exist in open-source ML-based systems?}

This research question seeks to uncover recurring logging smells that emerge during the development of ML systems. 
Inspired by prior empirical studies on software bad practices \cite{arnaoudova2013new, saarimaki2024taxonomy, zampetti2020empirical}, we aim to identify new concrete symptoms of poor or inconsistent logging practices in open-source ML projects hosted on GitHub. 
We formalize our research question in the following hypotheses:\\
\textbf{Hypothesis $H_1._1$:} ML codebases contain recurring logging smells.\\
\textbf{Hypothesis $H0_1._1$:} ML codebases do not contain recurring logging smells.\\ 

In this confirmatory design, these hypotheses will be evaluated through the systematic construction of a taxonomy: the presence of recurring patterns identified consistently across repositories constitutes evidence supporting $H_{1.1}$, whereas the absence of recurring or stable patterns would support $H0_{1.1}$. 
The resulting catalog will describe the types of logging smells, symptoms, and possible consequences of ML logging smells, with an example of a smell.

\subsection*{RQ2: How do ML practitioners perceive and experience these logging smells in practice?}

The second research question aims to evaluate how practitioners perceive the importance and impact of the logging smells identified in RQ1.\\ 
\textbf{Hypothesis $H_1._2$:} Practitioners recognize and rate logging smells.\\  
\textbf{Hypothesis $H0_1._2$:} Practitioners do not recognize and do not rate logging smells.\\  
This practitioner's perspective will allow us to validate the prevalence, relevance, and severity of the identified smells.
In this confirmatory design, these hypotheses will be evaluated through the analysis of survey responses: evidence for $H_{1.2}$ is provided when practitioners meaningfully differentiate smells through non-uniform Likert-scale ratings and demonstrate agreement patterns across participants, whereas uniformly neutral or inconsistent ratings would support $H0_{1.2}$.
Together, these research questions aim to produce:
\begin{itemize}
    \item an empirically grounded catalog of ML logging smells derived from open-source repositories, and
    \item an assessment of how practitioners rate the frequency and impact of these smells in their day-to-day ML development work. 
\end{itemize}

This dual perspective will bridge the gap between empirical evidence and practitioner insight, ultimately informing better logging guidelines and observability tools for ML development.

\section{Research Protocol}
\label{sec:RP}
This section details the research protocol that will be followed to answer the two research questions defined in this study, along with the variables that will be measured or controlled. 

\subsection{Variables}
\textbf{Independent Variable:} Type of Logging Library
\begin{itemize}
    \item \textbf{Description}: This variable distinguishes whether the project or code fragment primarily uses a \textit{general-purpose logging library} or an \textit{ML-specific logging framework}
    \item \textbf{Scale:} Categorical (general-purpose / ML-specific / hybrid).
\end{itemize}

\textbf{Dependent Variables:} Frequency of smells
\begin{itemize}
    \item \textbf{Description}: It represents how often a given logging anti-pattern appears in our sample data.
    \item \textbf{Scale:} Ratio.
    \item \textbf{Operationalization:} Computed automatically during taxonomy construction.
\end{itemize}


\textbf{Dependent Variables:} Perceived Severity.
\begin{itemize}
    \item \textbf{Description}: It captures how detrimental practitioners consider a given anti-pattern to be for ML development.
    \item \textbf{Scale:} Ordinal (5-point Likert).
    \item \textbf{Operationalization:} Mean response per pattern.
\end{itemize}

\textbf{Dependent Variables:} Perceived Frequency (Popularity).
\begin{itemize}
    \item \textbf{Description}: It measures how frequently practitioners observe each anti-pattern in their projects.
    \item \textbf{Scale:} Ordinal (5-point Likert).
    \item \textbf{Operationalization:} Mean response per pattern.
\end{itemize}

\subsection{Materials and Tasks}
This section describes the materials, data sources (RQ1), and tasks that survey participants will accomplish as part of the second phase (RQ2) of this study. 
All study materials, including code, datasets, and survey instruments, will be made publicly available online once finalized to ensure transparency and reproducibility.

\subsubsection{Materials}
Regarding the sample of repositories requested for conducting the empirical study (RQ1), we rely on the curated dataset and analytical infrastructure established in our prior empirical investigation of logging practices in ML-based applications \cite{foalem2024studying}. 
The materials used are summarized as follows:

\begin{enumerate}

    \item \textbf{Dataset of ML-Based Projects:} We will reuse the dataset of 502 open-source machine learning (ML) projects collected from GitHub (RQ1), as detailed in the previous study \cite{foalem2024studying}. 
    These projects were originally selected through a systematic search combining GitHub's API, ML-specific logging import configuration, and the Python programming language. 
    To ensure representativeness, repositories were included only if they satisfied minimum activity (at least 100 commits), popularity (minimum two contributors and two stars), and verifiable ML components (e.g., dependence on TensorFlow, PyTorch, scikit-learn).
    To maintain dataset relevance, we revalidated each repository's current status and removed archived or inactive projects, yielding a final set of 444 actively maintained ML-based repositories. 
    The dataset spans a broad range of ML application domains--including computer vision, natural language processing, and reinforcement learning--and reflects the dominant Python-based ML ecosystem.
    In total, these repositories contain 86,143 extracted logging statements, offering a sufficiently diverse and empirically rich foundation for identifying both traditional and ML-specific logging smells.

    \item \textbf{Logging Libraries:} We will also reuse the catalog of 12 logging libraries also identified by Foalem et al.~\cite{foalem2024studying} .These tools cover two primary categories: (i) \textit{general-purpose logging libraries} (e.g., \texttt{logging}), and (ii) \textit{ML-specific logging frameworks} (e.g., MLflow, TensorBoard, Weights \& Biases). 
    
    \item \textbf{LLM-Assisted Analysis:} We will use the GPT-4o-mini variant of GPT-4 due to its significantly lower cost per token, API stability, consistent availability, and deterministic outputs--properties that make it suitable for large-scale, reproducible classification workflows, especially when paired with human validation practices shown to be effective in prior LLM-assisted empirical studies \cite{abbassi2025unveiling, verdet2025assessing}. 
    The model will be prompted using a predefined instruction template to generate structured outputs (e.g., anti-pattern category, rationale, confidence score), which will then be manually validated by researchers.
    
    \item \textbf{Automation Scripts:} We will employ Python scripts to automate data preprocessing, logging statement parsing, LLM-based classification, and aggregation of results into structured tabular datasets. 
    These scripts will also support survey data analysis by automating the computation of descriptive and inferential statistics. The full replication package will be available here \cite{replication}.
    
    \item \textbf{Survey Instrument:} Regarding the RQ2, the proposed survey will be implemented using Google Forms to facilitate participation and ease of access. The form will contain structured items (Likert-scale questions) as well as open-ended prompts, enabling both quantitative and qualitative data collection from ML practitioners.
\end{enumerate}

\subsubsection{Tasks}
Specifically about the RQ2, we ask participants to validate the catalog of logging smells identified in the mining phase (RQ1) by evaluating their perceived relevance, frequency, and severity in real-world ML development contexts. 
Participants will complete the following tasks:

\begin{enumerate}[label=(\alph*)]
    \item \textbf{Validation of smells:} 
    Participants will be asked to review each anti-pattern category identified in RQ1. 
    For each category, We will provide the following information:
    \begin{itemize}
        \item A concise textual definition of the anti-pattern, summarizing its main characteristics and intent. 
        \item A representative code snippet extracted from real open-source ML projects to illustrate the anti-pattern in context.
    \end{itemize}
    This information will serve as a reference to help participants accurately evaluate each anti-pattern during the survey.
    For each subcategory, participants will rate their assessment on three 5-point Likert scales:
    \begin{itemize}
        \item \textit{Frequency (Popularity):} How frequently they encounter the target anti-pattern in their ML projects.
        \item \textit{Relevance:} How important they consider addressing this anti-pattern. 
        \item \textit{Perceived Severity:} How detrimental they believe the target anti-pattern represents to the ML development (e.g., reproducibility, maintainability, or observability).
    \end{itemize}

    \item \textbf{Proposing New smells:} 
    Participants will have the opportunity to suggest new smells or categories not covered in the provided catalog in a short comment, followed by a brief description and code snippet, if available.
    
    \item \textbf{Open Feedback:} 
    An open-ended section will be included for qualitative comments, enabling participants to share additional insights about their perceptions and opinions. 
\end{enumerate}
All responses will be anonymized, and participation will remain voluntary. 

\subsection{Participants}
This section describes the participants targeted for the survey phase of this study (RQ2), which aims to assess the perceived relevance, severity, and frequency of logging smells in machine learning (ML) development. 

\subsubsection{Target Participants and Recruitment Strategy}
Participants for this study will be \textit{ML practitioners}, including ML and MLOps engineers, and Data scientists and engineers. These participants will be recruited through two primary channels. 
First, open-source contributors will be contacted directly via the public email addresses they used to contribute to ML-related projects on GitHub, specifically those included in the dataset used in prior studies \cite{foalem2024studying, foalem2025logging}.
To avoid the perception of spam, we will design the recruitment email following the communication practices reported in Foalem et al.~\cite{foalem2025logging} and will conduct a preliminary test to ensure that the email is not flagged as spam before initiating full-scale contact.
These contributors have demonstrated practical experience with implementing and configuring logging mechanisms within ML pipelines. 
Second, to broaden participation beyond open-source contributors, the survey will also be publicly distributed through professional networks, including LinkedIn, to reach ML professionals working in industrial or research environments. 
The survey will remain open for one month: invitations will be sent at the beginning of the month, a reminder will be issued midway, and the survey will close at the end of the month. 
A similar recruitment and timing strategy has been effectively employed in prior studies \cite{abbassi2025unveiling,li2020qualitative}.
Before launching the full-scale survey, a pilot study involving 5-8 participants will be conducted to validate the clarity, readability, and relevance of the survey items. If the pilot reveals skewed participation--particularly underrepresentation of experienced practitioners--we will introduce participant compensation in the second recruitment wave.

\subsubsection{Ethical Considerations}
This study was approved by the Ethics Committee of Polytechnique Montreal (CER-2324-25-D). All participants will be informed of the study's purpose, the voluntary nature of their participation, and their right to withdraw at any time without consequence. 
No personally identifiable information will be published. 
The survey will not collect sensitive data beyond professional background and experience level. 
Ethical approval will be sought from our host institution's research ethics board prior to participant recruitment.

\subsection{Execution Plan}

This study will be executed in two major phases: (i) empirical mining and LLM-assisted analysis, and (ii) human validation and practitioner survey. 
The methodology builds upon the workflow proposed by Abbassi et al.~\cite{abbassi2025unveiling}, who systematically identified inefficiencies in LLM-generated Python code using a hybrid process that integrates automated judgments from LLMs with manual labeling and inter-rater validation. 

\subsubsection{Phase 1: Empirical Mining and LLM-Assisted Identification of Logging smells}

The first phase aims to empirically mine ML-based open-source repositories to identify recurring logging smells. 
For that, as previously discussed, we plan to reuse the curated dataset of 444 ML-based projects collected in previous work~\cite{foalem2024studying}. 

\paragraph{Step 1: Data Verification and Extraction}
We will begin by generating an Abstract Syntax Tree (AST) for the files, and then extract all Python functions that contain at least one logging statement associated with the previously selected tools \cite{foalem2024studying}. 
Files that include multiple functions with logging statements will be grouped to allow for contextual and file-level analysis of logging behaviors. 
Each extracted function and file will be stored with metadata, including repository name, library type (general-purpose or ML-specific), and file path. Our replication package contains all the scripts we started developing and the final version of our dataset in Json format\cite{replication}.


\paragraph{Step 2: LLM-Assisted Identification of smells}
Following the methodology proposed by Abbassi et al.~\cite{abbassi2025unveiling}, we will first construct a representative statistical sample of Python files containing logging statements. Using Yamane's formula for sample size estimation with a 95\% confidence interval~\cite{yamane1973statistics}, we will randomly select a subset of files from our dataset to ensure statistical representativeness. 
The sampling will be stratified according to the type of logging library used in each file--\textit{ML-specific}, \textit{general-purpose}, or \textit{hybrid} (files containing both)--to preserve proportional representation across library categories. 
This stratification also helps reduce the overall manual analysis effort by ensuring that all logging categories are proportionally represented in the sample, avoiding over-analysis of dominant categories and under-analysis of less frequent ones.
Once the sample is established, we will employ GPT-4o-mini to perform an initial classification of potential logging smells. 
The model will be prompted using a structured template that provides code snippets and concise task instructions, adopting the persona of an expert in ML observability and software quality.

The prompt will instruct the model to identify incorrect or suboptimal logging practices at both the \textit{function level} and the \textit{file level}. 
Each model output will be in JSON format, including a short explanation and a proposed anti-pattern label. 
All generated outputs will be recorded and systematically linked to their corresponding code fragments for traceability and later validation.

\paragraph{Step 3: Human Validation and Iterative Annotation}
We will conduct an iterative, LLM-assisted annotation process to ensure both coverage and reliability of identified smells. 
The analysis will proceed in cycles of 100 files, following a progressive refinement strategy designed to gradually stabilize the taxonomy of logging smells.

\begin{itemize}
    \item \textbf{Initial Cycle (10\% + 40\% + 50\% Split):} In the first iteration, 10\% of the files will be used to prompt the LLM and generate an initial sample of candidate logging smells. 
    Two authors will manually review these outputs and develop preliminary definitions using an open coding approach~\cite{seaman1999qualitative} during a meeting discussion. 
    The refined taxonomy will then be provided to the LLM as contextual knowledge to classify an additional 40\% of the files with the possibility to suggest new anti-pattern categories. During a meeting discussion, two authors will review the evaluation reported by the LLM. 
    The remaining 50\% of files in this initial batch will be manually annotated by two researchers using the emerging taxonomy. 
    Inter-rater reliability will be computed using Cohen's $\kappa$ to assess consistency and identify ambiguous cases for refinement \cite{mchugh2012interrater}.

    \item \textbf{Subsequent Cycles (50\% LLM + 50\% Manual):} 
    In all following iterations, each new batch of 100 files will be divided equally: 50\% will be analyzed using LLM-assisted classification, and the remaining 50\% will be manually annotated by two researchers. 
    The LLM will be instructed to map each snippet to the existing taxonomy while suggesting new categories only if it identifies behaviors not previously defined. 
    Human reviewers will validate both the LLM classifications and any newly suggested smells to maintain taxonomic coherence. 
    For each cycle, Cohen's $\kappa$ will be recalculated to monitor agreement between annotators and to ensure reliability over time.
\end{itemize}

This iterative procedure will continue until \textit{theoretical saturation} is achieved--that is, no new anti-pattern categories emerge, and existing classifications remain stable across consecutive cycles. 
This approach ensures continuous validation between LLM-generated classifications and human expert judgment, since LLMs can hallucinate~\cite{sollenberger2024llm4vv}. 

\paragraph{Step 4: Taxonomy Construction}
Once saturation is achieved, a comprehensive taxonomy of ML logging smells will be synthesized. 
Each taxonomy entry will include: 
(i) a concise definition, 
(ii) a representative code snippet, and 
(iii) the observed frequency of occurrence across projects. 
This final taxonomy will serve as the foundation for the practitioner survey in Phase~2, enabling the assessment of perceived severity, practical relevance, and popularity of each identified anti-pattern.

\subsubsection{Phase 2: Practitioner Survey and Validation}

In the second phase, we will conduct a structured online survey to validate the catalog of logging smells identified in Phase~1 and to assess their perceived relevance, frequency, and severity from the perspective of ML practitioners. 
The survey will be administered using Google Forms \footnote{\url{https://docs.google.com/forms/u/0/}}.
The survey form will consist of the following structured sections:

\begin{itemize}
    \item \textbf{Welcome Page:} 
    The opening section provides a brief introduction to the study, its objectives, and the expected contributions to enhancing logging practices in ML-based software systems. 
    This page will also include the name, institutional affiliation, and contact information of the project lead to ensure transparency and enable participants to ask questions, if necessary.
    At the end of this page, participants will be asked for their explicit consent to participate in the survey through a yes/no confirmation question.

    \item \textbf{Demographic Information:} 
    This section will collect information about the participants' professional background and experience to characterize the diversity of the respondent pool. 
    Questions will cover the following aspects:
    \begin{itemize}
        \item Current company or organization name (optional);
        \item Current professional position or role (e.g., ML engineer, data scientist, software developer);
        \item Years of experience in software engineering, Python programming, and machine learning development.
        \item A brief skills assessment consisting of 2--3 multiple--choice questions evaluating familiarity with logging practices and ML development workflows (e.g., identifying proper logging usage or recognizing ML-specific logging tasks).
    \end{itemize}
    These data will allow us to analyze responses across experience levels and professional contexts.

    \item \textbf{Logging Practices:} 
    Participants will first be asked whether they use logging in ML-based applications (\textit{yes/no}). 
    If the response is \textit{yes}, they will proceed to the main validation section; otherwise, the survey will conclude after a brief thank-you message.

    \item \textbf{Validation of Logging smells:} 
    For participants who report using logging, we will first ask about the tools they have used before. Then, we will present the catalog of logging smells identified in Phase~1. 
    Each anti-pattern will include:
    \begin{itemize}
        \item A concise textual description of the anti-pattern;
        \item A representative Python code snippet extracted from open-source ML projects to illustrate its context.
    \end{itemize}
    For each anti-pattern, participants will be asked to rate three aspects on 5-point Likert scales:
    \begin{itemize}
        \item \textit{Relevance:} how important they consider addressing the target anti-pattern in practice;
        \item \textit{Popularity:} how often they encounter the given anti-pattern in their own ML development experience;
        \item \textit{Severity:} how detrimental they perceive the target anti-pattern to be for maintainability, observability, or reproducibility of ML systems.
    \end{itemize}
    Finally, an open-ended text box will also be provided after each section to allow participants to share additional comments, and a final text box to describe smells not captured in the catalog.
\end{itemize}
The survey is designed to require approximately 15-20 minutes to complete. 
All responses will be collected anonymously, and participation will remain entirely voluntary. 
The resulting data will be used to evaluate the practical significance of each logging anti-pattern and to refine the taxonomy developed in Phase~1 based on real-world practitioner feedback.

\subsection{Analysis Plan}

To address \textbf{RQ1}, we will employ a mixed-method approach combining qualitative and quantitative analyses to construct the taxonomy of ML logging smells. 
The qualitative component will rely on the identification and categorization of logging smells, as described in the previous section. 
The quantitative component will measure the frequency of each identified anti-pattern across the analyzed projects, allowing us to determine how commonly these patterns occur in ML-based repositories.

For \textbf{RQ2}, we will apply a similar mixed-method strategy to analyze the survey responses collected from ML practitioners. 
The quantitative analysis will summarize Likert-scale ratings on the perceived \textit{relevance}, \textit{frequency}, and \textit{severity} of each anti-pattern using descriptive statistics and comparisons across participant demographics. 
The qualitative analysis will involve thematic analysis of open-ended responses to uncover additional insights or new smells not represented in the initial taxonomy.





\section{THREATS TO VALIDITY}
\label{sec:threats_to_validity}
This section discusses potential threats to the validity of the study and the strategies we will adopt to mitigate them.

\paragraph{Construct Validity} 
The threats to the construct validity in our study are related to whether the study variables and measures accurately represent the underlying concepts. 
For RQ1, the identification and classification of ``logging smells'' may introduce interpretation bias during annotation. 
To mitigate this, we will employ a systematic open coding procedure supported by LLM-assisted classification, followed by independent validation by two researchers. 
Inter-rater agreement will be measured using Cohen's $\kappa$ to ensure consistency in categorization. 
For RQ2, potential response bias or misunderstanding of survey items may occur due to participants' differing interpretations of smells. 
To address this, the survey will include clear definitions and representative code snippets for each anti-pattern, aiming to ensure a consistent understanding among all respondents.

\paragraph{Internal Validity} 
A key internal threat to validity is the presence of confounding factors that could influence the results of our analyses. 
For RQ1, the automated extraction and LLM-based classification of logging statements may produce false positives or false negatives, potentially misrepresenting the true distribution of smells. 
To mitigate this risk, we will manually validate a statistically significant subset of LLM outputs, iteratively refine prompt instructions, and adjust annotation criteria until consistent classification performance is achieved. 
For RQ2, participant self-selection bias may occur if certain practitioner profiles--such as those with greater ML experience or stronger opinions on logging--are more likely to respond. 
We will address this by recruiting participants from diverse sources, including open-source contributors and industry professionals, and by reporting the demographic distribution of respondents to ensure transparency and contextualize the results.

\paragraph{External Validity} 
One external threat to validity concerns the representativeness of our dataset and the extent to which our findings can be generalized beyond the analyzed sample. 
Our dataset, which includes 444 recent and active open-source Python-based ML projects, provides a strong empirical foundation but may limit generalizability to other programming languages or proprietary industrial systems. 
Similarly, the analysis relies on a predefined set of logging libraries identified in prior work, based on a dataset released within the last two years. 
While this recency ensures that our sample reflects current ML development practices, it may not fully capture emerging or domain-specific logging frameworks that have appeared more recently. 
To mitigate this limitation, we complement the repository mining with a practitioner survey (RQ2) to incorporate perspectives from both open-source and industry practitioners. 
As future work, the dataset can be extended to include newer or evolving logging libraries as well as additional programming languages (e.g., Java, C++, R) to evaluate whether the identified smells persist across different technological ecosystems and development contexts.

\paragraph{Conclusion Validity}
Conclusion validity concerns whether the procedures used in this study allow for accurate and justifiable inferences about the presence and characteristics of logging smells in ML-based systems. To strengthen conclusion validity, we rely on a confirmatory design that specifies all analysis procedures a priori, including sampling, coding, and survey evaluation. The iterative LLM-assisted coding process is always validated by human coders, with inter-rater agreement (Cohen's $\kappa$) used to assess the reliability of classifications. The use of stratified sampling ensures that conclusions are not disproportionately influenced by overrepresented logging library types. For the practitioner survey, multiple complementary measures (perceived frequency and severity) reduce the risk of relying on a single subjective indicator. Together, these methodological safeguards help ensure that the relationships and patterns identified in the study are robust, interpretable.


\bibliographystyle{IEEEtran}
\bibliography{Paper.bib}

\end{document}